\documentclass[twocolumn,dvipsnames]{aastex631}
\usepackage{ApJLPreamble}
\graphicspath{{./figures/}}
\newcommand{\new}[1]{{\color{blue}#1}}

\newcommand{\CUHK}{
Department of Physics, The Chinese University of Hong Kong, Shatin, N.T., Hong Kong
}
\newcommand{\ICTS}{
International Centre for Theoretical Sciences, Tata Institute of Fundamental Research, Bangalore 560089, India
}
\newcommand{\CIFAR}{
Canadian Institute for Advanced Research, CIFAR Azrieli Global Scholar,
MaRS Centre, West Tower, 661 University Ave, Toronto, ON M5G 1M1, Canada
}
\newcommand{\ie}{{\it i.e.}}
\newcommand{\eg}{{\it e.g.}}
\newcommand{\cf}{{\it c.f.}}

\def\<#1>{\mathinner{\langle#1\rangle}}

\usepackage{layouts}
\usepackage{csquotes}
\renewcommand{\new}[1]{#1}

\begin{document}
\title{Constraining binary mergers in AGN disks using the non-observation of lensed gravitational waves}

\author[0000-0003-0470-282X]{Samson H.\,W. Leong}
    \email{samson.leong@link.cuhk.edu.hk}
    \affiliation{\CUHK}
    
\author[0000-0003-2888-7152]{Justin Janquart}
    \affiliation{Department of Physics, Utrecht University, Princetonplein 1, 3584 CC Utrecht, The Netherlands}
    \affiliation{Nikhef, Science Park 105, 1098 XG Amsterdam, The Netherlands}
    \affiliation{Center for Cosmology, Particle Physics and Phenomenology - CP3, Universit\'e Catholique de Louvain, Louvain-La-Neuve, B-1348, Belgium}
    \affiliation{Royal Observatory of Belgium, Avenue Circulaire, 3, 1180 Uccle, Belgium}

 \author[0000-0003-0067-346X]{Aditya Kumar Sharma}
    \affiliation{\ICTS}

\author[0000-0001-9765-7735]{Paul Martens}
    \affiliation{\CUHK}
    
 \author[0000-0001-7519-2439]{Parameswaran Ajith}
    \affiliation{\ICTS}
    \affiliation{\CIFAR}

 \author[0000-0002-3887-7137]{Otto A. Hannuksela}
    \affiliation{\CUHK}
\begin{abstract}
    \noindent
    The dense and dynamic environments within active galactic nuclei (AGN) accretion disks may serve as prolific birthplaces for binary black holes (BBHs) and one possible origin for some of the BBHs 
    detected by gravitational-wave (GW) observatories. 
    We show that a considerable fraction of the BBH in AGN disks will be strongly lensed by the central supermassive black hole (SMBH). Thus, the non-observation of lensed GW signals can be used to constrain the fraction of BBH binaries residing in AGN disks.
    The non-detection of lensing with current ${\cal O}(100)$ detections will be sufficient to start placing constraints on the fraction of BBHs living within accretion disks near the SMBH. In the next-generation detectors era, with ${\cal O}(10^5)$ BBH observations and no lensed events, we will be able to rule out most migration traps as dominant birthplaces of BBH mergers; moreover, we will be able to constrain the minimum size of the accretion disk. 
    On the other hand, should AGNs constitute a major formation channel, lensed events from AGNs will become prominent in the future.
\end{abstract}
\section{Introduction}
Active galactic nuclei (AGN) are promising sites for stellar-origin black hole (BH) mergers~\citep{Ford2019:AGNWhitePaper,Ford2022:MergerRate}. 
These binary black holes (BBHs) reside in the accretion disk of the AGN, orbiting the centre supermassive black hole (SMBH), and merge within a rapid timescale~\citep{Secunda2019:BBHFormation}. The resulting gravitational waves (GWs)~\citep{Ishibashi2020:GW} could be detected by current and future ground-based  detectors~\citep{AdvLIGO,TheVirgo:2014hva,KAGRA,US_CE,Grado2023:ET,NEMO}.
These mergers are expected to be particularly abundant near the migration traps -- regions in the disk where inward and outward torques exerted by the gas disk cancel~\citep{Yang2019:MassDist,McKernan2012:IMBHFormation,Thompson2005:AGNdisk,Peng2021:LastMigration}. The detection of such BBHs may provide information about their AGN environment~\citep{Wang2021:AGNBBH_pop,McKernan2020:MCMC_BBHAGN,Stone2016:AGNBBH_hardening}, such as mass of the SMBH and the age of the AGN~\citep{Vajpeyi2022:AGNProperties}.
Remnant BHs from such mergers, rapidly moving through the gas disk due to the GW recoil, could produce a transient electromagnetic flare~\citep{Graham:21gZTF,Graham2023:GWTC3,McKernan2019:flare,Wang2021:AccretionJet,Leob2007:flare}. This could be observable for an observer perpendicular to the disk (face-on). When viewed from the plane of the disk (edge-on), the flare would be blocked by the gaseous torus surrounding the disk. However, if the BBH is located behind the SMBH, then the GW signal can be strongly lensed by the SMBH, splitting it into two \enquote{images} of the same GW~\citep{Takahashi2003:WaveOptics,Millon2023:AGNlensing}.
These images would appear as repeated waves with similar properties, except for differing amplitudes, arrival times, and phases.
These lensed signals can be detected using standard lensing search methods~\citep{Haris2018:IdentifySL,Liu2023:Millilensing}. 
In this letter, we show that a fraction of such mergers produce detectable lensed GWs, with a lensing probability that depends on their radial distribution on the accretion disk.
Consequently, the non-detection of lensed GW events will constrain the number of such systems and exclude regions of the disk as the dominant BBH merger sites.
\begin{figure}[ht]
    \centering
    \includegraphics[width=.485\textwidth]{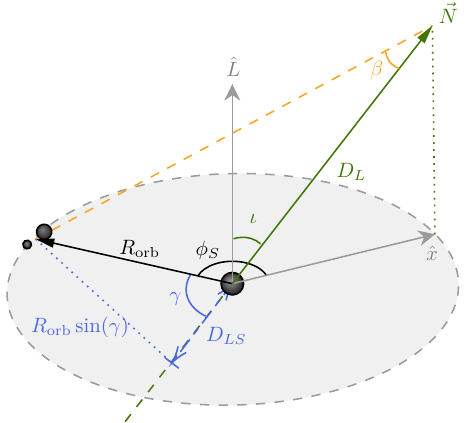}
    \caption{
        {\bf Schematic diagram of the AGN-BBH system.} 
    Diagram showing a BBH (left) orbiting an SMBH (center).
    The grey ellipse is the orbital plane of the BBH around the SMBH, defined by the orbital angular momentum $\vu*L$.
    The accretion disk is modelled as a thin disk, coplanar with the orbital plane.
    The observer is located at $\va*N$ and represented by the green vector; its continuation below the plane is the green dashed line, and they form the optical axis.
    The inclination angle is $\iota$, and the projection of $\va*N$ onto the orbital plane defines the $x$-axis, from which the azimuthal angle of the BBH, $\phi_S$, is measured.
    The angles subtended by the source and the SMBH, and the source and the observer are denoted by $\gamma$, and $\beta$, respectively. 
}
    \label{fig:agn_BBH_illustration}
\end{figure}
\begin{figure}[ht]
    \centering
    \includegraphics[width=.5\textwidth]{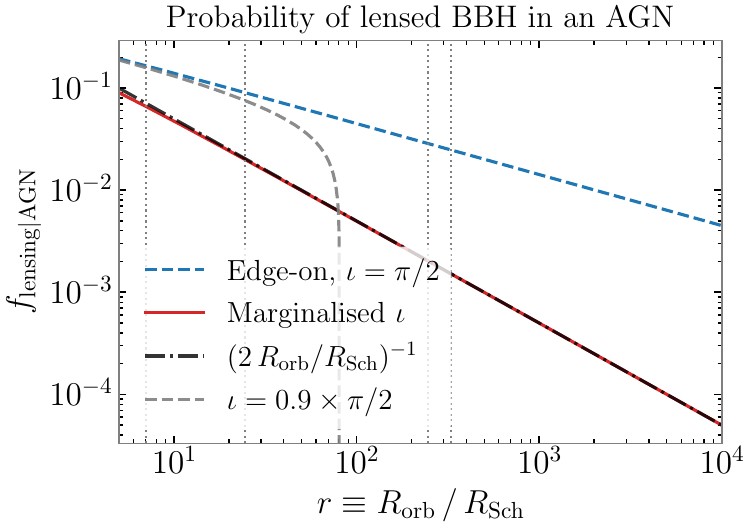}
    \caption{
    {\bf BBH lensing probability as a function of $R_{\rm orb}$.}
    The dashed blue and grey lines represent the lensing probability (Eq.~\eqref{eq:conditional_prob}) at fixed inclinations $\iota = \pi/2$ and $9 \pi / 20$, respectively. 
    The former follows the $1 / \sqrt{r}$ trend, and the latter has a finite radius cut-off, at around $80\,R_{\rm Sch}$. 
    Shown in red is the marginalised probability over inclination (Eq.~\eqref{eq:marg_prob}, or \eqref{eq:exact_f_lensing_AGN}), which matches very closely $1 / (2 r)$ (black dash-dotted line, Eq.~\eqref{eq:prob_approx}), except for at small radius (for $r \lesssim 10$).
    Vertical dotted lines indicate the locations of some migration traps mentioned in the literature (\eg ~\citet{Bellovary2016:Migration,Peng2021:LastMigration,Thompson2005:AGNdisk,Sirko2003:AGNdisk}), where a majority of binary black holes are expected to migrate to and merge.
    }
    \label{fig:lensed_prob}
\end{figure}
\section{The AGN-BBH system\label{sec:AGN-BBH}}
We consider a BBH residing within the accretion disk of an AGN, and orbiting around a SMBH of mass $M_{\rm SMBH}$ at a fixed radius $R_{\rm orb}$. We refer to this kind of binary as AGN-BBH. 
We define the $x$-axis by the projection of the observer onto the orbital plane, and $\phi_S$ is the azimuthal angle of the BBH with respect to that $x$-axis (Fig.~\ref{fig:agn_BBH_illustration}).
For an observer at an angle $\iota$, the probability of observing a binary that is located at a radius\footnote{The Schwarschild radius is $R_{\rm Sch} = 2 G M_{\rm SMBH} / c^2$.} $r = R_{\rm orb} / R_{\rm Sch}$ as strongly lensed (\ie\ the binary falls within the volume defined by the Einstein radius $\theta_{\rm E}$, Eq.~\eqref{eq:beta_bound}), is given by:
\begin{equation}
    p({\rm lensing} \mid r,\ \iota) = \frac{1}{\pi}\arccos( \frac{x}{\sin\iota})\ ,
    \label{eq:conditional_prob}
\end{equation}
where $x$ is defined as
 \begin{equation}
     x = \frac{\sqrt{1 + r^2} - 1}{r}\ .
     \label{eq:x_def}
\end{equation}
Assuming that observers are distributed isotropically around the SMBH (\ie, no preferred orientation for AGNs), we can marginalise over $\iota$ to compute the lensing probability for BBH \emph{located at a radius $r$} in the AGN disk
\begin{align}
    f_{{\rm lensing}|{\rm AGN}} &= p({\rm lensing} \mid r) \\
    &= \int_{\arcsin{x}}^{\pi /2} \frac{1}{\pi}\arccos( \frac{x}{\sin\iota}) \, \sin(\iota)\,\dd  \iota  \label{eq:marg_prob}\\
    &\approx \frac{1}{2 \, r}\ . \label{eq:prob_approx}
\end{align}
The final approximation is only valid for large $r$ (for $r \gtrsim 10$, see Fig.~\ref{fig:lensed_prob}) and the full derivation can be found in the appendix.
\section{Constraining the fraction of BBHs in AGN disks}
Knowing the probability of observing a lensed BBH in an AGN disk, we show that the non-detection of lensed GW events can then be used to constrain the fraction of BBHs that reside in AGN disks
($f_{\rm AGN}$) among all BBH detections.
First, we define $f_{\rm AGN}$ as the fraction of GW events that are originated from AGN among all $N_{\rm obs}$ \emph{observed} GW events:
\begin{equation}
    f_{\rm AGN} \equiv \frac{N_\text{obs. BBH from an AGN}}{N_{\rm obs.}}\ ,
\end{equation}
then, the probability of observing a lensed BBH from AGN is:
\begin{equation}
    f_{{\rm lensed} \wedge {\rm AGN}}  = f_{\rm lensing|AGN} \times f_{\rm AGN}\ ,
    \label{eq:lensed_and_AGN}
\end{equation}
where the conditional probability characterises the probability of detecting a lensed binary from an AGN, which is subject to the AGN geometry and potential detection loss due to demagnified images. We will address the latter issue in the discussion and appendix. For the remainder of this paper, we will assume each lensed AGN-BBH will be correctly detected and identified.
For a given number of GW observations and $f_{\rm AGN}$, this probability can be used to estimate the expected number of lensed events via the binomial distribution.

A non-detection of lensed events can, therefore, be used to constrain the fraction of AGN events among all observed binaries.
Following~\citet{Basak2022:MACHOS}, we define the {upper bound} of this AGN fraction to be the 90\textsuperscript{th} percentile of the conditional distribution of $f_{\rm AGN}$, \ie\ the $f_{\rm AGN}^{\rm upper}$ satisfies:
\begin{equation} 
    90\% = \int_0^{f_{\rm AGN}^{\rm upper}} \,\dd f_{\rm AGN}\, p(f_{\rm AGN} \mid N_{\rm obs}, N_{{\rm lensed}}=0)\ . 
    \label{eq:fraction}
\end{equation}

\begin{figure*}[!ht]
    \centering
    \includegraphics[width=.49\textwidth]{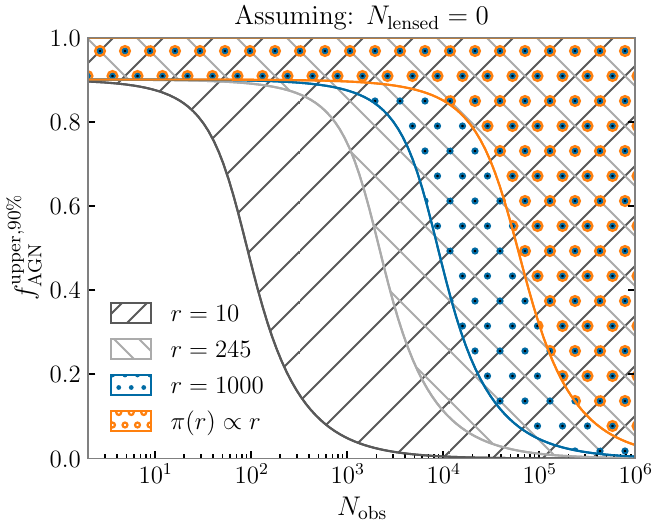}
    \includegraphics[width=.49\textwidth]{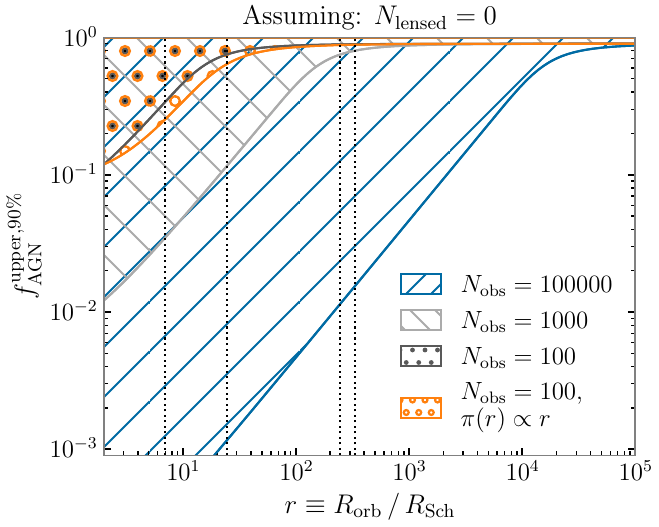}
    \caption{
        {\bf Constraints on the fraction of AGN-BBH events based on the non-detection of lensed events.}
        The left and right panels show the 90\textsuperscript{th} percentile upper bound of $f_{\rm AGN}$ as a function of the number observed BBH events $N_{\rm obs}$ and the radial distance $r$ from the SMBH, respectively.
        The hatched area represents the $f_{\rm AGN}$ that is ruled out by the non-detection of lensed events.
        In the left panel, each curve corresponds to a different fixed radial distribution of BBH around the SMBH; in particular, two grey and blue curves (hatched from the top-right and the top-left, and small circles) represent BBHs from fixed radii $r =$ 10, 245 and 1000, respectively. The orange (circles) curve accounts for the whole disk case by assuming the number of BBHs grows linearly with the radius (\ie\ BBHs distributed uniformly on the disk, with $2 \leqslant r \leqslant 10^4$). 
        The right panel shows the disk inner region that will be excluded as the dominant BBHs production site, for three different numbers of observed events, as well as for the number of mergers growing linearly at $N_{\rm obs}=100$ (orange, circles). 
        Note that in the last case, the $x$-axis represents the maximum disk radius, $2 \leqslant r \leqslant  r_{\rm max}$, with the same meaning as in Fig.~\ref{fig:f_AGN_disk_constraint}.
        It is worth noting that, following from the definition Eq.~\eqref{eq:fraction}, $f_{\rm AGN}^{{\rm upper}, 90\%}$ is 0.9 for regions with no constraint.
        Finally, the vertical dotted lines denote the trap locations reported in~\citet{Bellovary2016:Migration} and \citet{Peng2021:LastMigration}, at $r \in \{7,\ 24.5,\ 245,\ 331\}  $ respectively.
    }
    \label{fig:f_AGN_constraint}
\end{figure*}

Figure~\ref{fig:f_AGN_constraint} shows the constraints on $f_{\rm AGN}$ if we do not detect any lensed events, assuming different numbers of observed events, and across a range of orbital radii around the SMBH.
With the ${\cal O}(100)$ events from the LIGO-Virgo-KAGRA (LVK) third observation run (O3)~\citep{GWTC-2.1:catalog}, from the left panel, we find that \emph{if} all AGN-BBHs merge near the \enquote{last migration trap} (${\sim}10\,R_{\rm Sch}$)~\citep{Peng2021:LastMigration}, then the fraction of AGN-BBHs events cannot be over 47\% of all LVK binaries. 
This will be limited to within 5\% when we have ${\cal O}(1000)$ (unlensed) observations. 
Most other migration traps reported are located between 100\,-\,1000\,$R_{\rm Sch}$~\citep{Bellovary2016:Migration,Grishin2023:thermalTrap}, as an example, the light grey (hatched from the top-left) curve represents the constraint for binaries at $245\,R_{\rm Sch}$. In this case, constraining the AGN-BBH fraction to less than 50\% will require ${\sim}2200$\ events.

We also show, in orange (circles), the constraint when the BBHs are distributed uniformly over the entire disk (number density growing linearly with the radius, with $2 \leqslant r \leqslant 10^4)$~\footnote{
In this simple model, with $\pi(r) \propto r$, the marginalised lensing probability is $\approx 1 / r_{\rm max}$. Thus, the lensing probability for this distribution is the same as that of the migration traps model, where the binaries are located at $r = r_{\rm max}/2$ (\cf\ Eq.~\eqref{eq:prob_approx}).}.
\begin{figure}[!h]
    \centering
    \includegraphics[width=.49\textwidth]{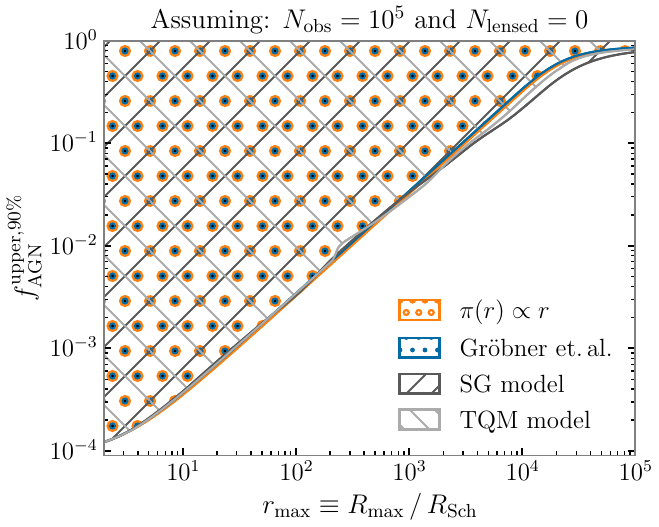}
    \includegraphics[width=.49\textwidth]{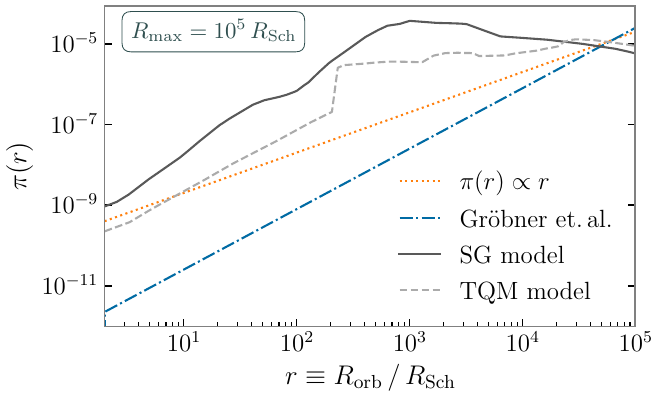}
    \caption{
        {\bf Constraints on the fraction of AGN-BBH events when they are distributed over the entire disk.} 
        The 90\textsuperscript{th} percentile upper bound of $f_{\rm AGN}$ is shown as a function of the maximum AGN disk radius, assuming four disk models and with $10^5$ observations, in the upper panel.
        The hatched area represents the $f_{\rm AGN}$ that is ruled out by the non-detection of lensed events.
        The orange curve denotes the same simple disk model as in Fig.~\ref{fig:f_AGN_constraint}.
        The blue curve assumes the BBHs follow the radial distribution given by Eq.~(19) of~\citet{grobner_binary_2020}.
        The last two, dark and light grey, have their radial distributions constructed from the density and height profiles from ~\citet{Sirko2003:AGNdisk} (SG) and ~\citet{Thompson2005:AGNdisk} (TQM), respectively. 
        In the bottom panel, we show the several radial distributions we employed, in log-log scale, normalised to $r_{\rm max} = 10^5$, following the same color codes as above.
    }
    \label{fig:f_AGN_disk_constraint}
\end{figure}
In the right panel of Fig.~\ref{fig:f_AGN_constraint}, we show the upper limit of $f_{\rm AGN}$ as a function of orbital radius instead. This is useful for understanding the constraints on different AGN disk models that predict AGN-BBHs located at a narrow range of radii. 
The dark grey curve (with small circles) is the constraint achieved with 100 events, which can only place meaning bounds for AGN-BBHs located within $100\,R_{\rm Sch}$ from the centre, and at most down to ${\sim} 12\%$. 
In the era of third-generation detectors when the number of BBH events are expected to reach $\mathcal{O}(10^5)$ per year~\citep{Evans2023:CE_prospect}, a non-detection of lensed events can limit the fraction of GW events from AGNs down to $0.5 - 5\%$ if the AGN-BBHs are located in the range $100 \leqslant r \leqslant 1000$,
where the migration traps are expected to form, denoted by vertical grey dotted lines. This will enable us to rule out these trap locations~\citep{Bellovary2016:Migration,Grishin2023:thermalTrap} as dominant AGN-BBH production sites.
We note that, however, there are possible trap locations at $r > 10^4$, as suggested in~\citet{Tagawa2020:CompactFormation}, which will require much longer observation time to be constrained.

Finally, Fig.~\ref{fig:f_AGN_disk_constraint} shows the $f_{\rm AGN}$ constraint
taking the contribution from the whole disk into account as a function of their maximum disk radii, $R_{\rm max}$, while assuming four different disk models with $10^5$ BBH observations. 
We find that other than small fluctuations at particular radii due to the different density profiles, the overall lensing probability drops with $r_{\rm max}$ regardless of models, and hence results in the largely similar constraint on $f_{\rm AGN}$. 
This result also suggests that the non-detection of lensing is most effective at informing the minimum size of the disks, while the effects from different disk models only become significant for $r_{\rm max} \gtrsim 2000$. 
\section{Discussion}
\citet{Graham:21gZTF} reported a plausible electromagnetic counterpart to the BBH merger event GW190521, which is consistent with a flare produced by the motion of the remnant BH through an AGN disk. Following this,~\citet{Graham2023:GWTC3} performed a search for electromagnetic counterparts to candidate BBH triggers from O3 and found at most 7 pairs of GW-EM signals\footnote{There are 12 event pairs listed in Table 3 of~\citet{Graham2023:GWTC3}; but we find at most 7 of those can be simultaneously true, for some of the 12 pairs share the same GW or EM signal.}, with an expectation of 3 pairs when $f_{\rm AGN} = 0.5$. 
If those were real, they imply the lower bound of $f_{\rm AGN} \approx 1$ (\cf~Eq.\,(14) thereof).
If we further assume all these binaries are GW190521-like, located in the migration trap at $r \approx 331$~\citep{Thompson2005:AGNdisk,Graham:21gZTF}, 
then we find that with ${\sim}100$~(1000) observations, the probability of observing at least one lensed event is 14\%~(78\%). 

Some limitations of this work are worth mentioning. 
For one, we used a thin disk model for the accretion disk.
This assumption could be refined, but even in a full disk model, the typical aspect ratio is less than 10\%.
Preliminary investigations do not show significant changes to our conclusions.

When the BBH is located at a large distance from the optical axis, the second lensed image will be demagnified. This will reduce the horizon distance to lensed AGN-BBH systems from which both images are detectable.
We have estimated the loss of binaries due to this is 60\% in the worst-case scenario. 
Furthermore, when the sensitivity of the detectors is higher (\eg, next-generation detectors), and the horizon distance extends far beyond where these binaries are expected to form, first, we find that effects due to demagnification will be negligible. 
Moreover, when that is the case, then there are no selection effects induced by the redshift distribution. 
In this case, the $f_{\rm AGN}$ that we will report is not limited to the \emph{observed} population of AGN-BBH, but the \emph{actual} fraction of binaries in AGN disks.
Note that even though there could be additional effects on the gravitational lensing of binaries in the vicinity of the SMBH, such as wave-optics effects at low GW frequencies \new{which will be relevant for future space-based GW detectors~\citep{Takahashi2003:WaveOptics,DOrazio2020:Hierarchical,Pijnenburg2024:LISA}}, or the potentially highly-magnified overlapping signals~\citep{RicoLo2024:HighMag}, they shall be addressed by future infrastructures for detection of lensed GWs, which shall not hamper the merit of this work as long as these lensed signals are correctly identified. 
\new{Nevertheless, if there are other stellar objects in the vicinity of the SMBH with much lower masses, then they could induce additional microlensed images, similar to the scenario considered by \citet{Mishra2021:StellarML}.}
Moreover, for sources at distances as close as ${\cal O}(10)\,R_{\rm Sch}$, \new{the spin nature of GW can induce geodesics that are frequency- and polarisation- dependent~\citep{Harte2022:SpinHall,Oancea2022:FreqPolarisation,Oancea2023:SpinOrbit}, effectively leading to birefringent effects to the lensed images.} Also, the thin-lens approximation might not hold strongly at this distance. To this end, we have checked that changes in the deflection angle are sufficiently small at $r = 10$ using the exact lens equation~\citep{Bozza2008:LensEqn}. 
Nevertheless, this part of the disk will only contribute to a small proportion of the whole disk, and such deviation is negligible.
Finally, while we have here excluded event counts from binary neutron stars (BNSs), the accretion disk could also host BNSs, the merger rate of which is correlated with that of BBHs~\citep{McKernan2020:MergerRate}.
For the next-generation detectors, where the expected BNS detections are higher than that of BBH, this shall render the prospective constraints even stronger. 

\section{Conclusion}
In this work, we showed that a considerable fraction of the BBH in AGN disks will be strongly lensed by the central SMBH. Thus, the non-observation of lensed GW signals can be used to constrain the fraction of BBH binaries residing in AGN disks. While the current constraints are modest,
as the number of observed GW signals increases, this will allow us to set tight constraints on the fraction of AGN-BBH systems among all detections. 
In particular, we will be able to rule out some locations of migration traps as the dominant BBH merger sites, as well to put a lower bound on the size of the accretion disk. 
Future GW observatories that are expected to detect ${\cal O}(10^5)$ events
per year, should tighten these bounds further to the sub-percent level within their first year of operation. 
On the other hand, if AGN-BBH is a major formation channel for BBHs, then we can expect to observe a sizable number of lensed events in the future.
Previous studies based on the existence of AGN-BBH systems~\citep{Ford2022:MergerRate,Graham2023:GWTC3} and spatial correlation~\citep{veronesi_detectability_2022,bartos_gravitational-wave_2017} placed some lower bounds on the fraction of AGN-BBH events. 
Here, we present the upper-bound of $f_{\rm AGN}$ based on the non-detection of lensed events, which will further allow us to sandwich the possible $f_{\rm AGN}$ range and constrain different AGN models.
Besides evaluating the spatial correlation of the electromagnetic counterpart of AGN-BBH GW signals, confirmation of such AGN-BBH events can also rely on measuring the recoil direction~\citep{Leong2024:AGNKick,Bustillo:GW190412}, and the deviations in waveform morphology due to the orbital motion around the SMBH~\citep{Morton2023:21gAGN,Vijaykumar2023:LOSAcc,Vijaykumar2022:BNS,TorresOrjuela2021:MovingExcitation}, the dense gaseous environment, which particularly affects the ringdown signal~\citep{Leong2023:ScalarField}. 
As a final remark, we note that confirming and characterising such AGN-BBH events is also possible via the characteristic lensing signatures from such systems~\citep{Gondan2021:GNLensing}, or potentially the large opening angle between the two lensed GW images and Doppler lensing, which we shall defer to future work to explore. 
\begin{acknowledgments}
We would like to thank Barry~McKernan and Saavik~Ford for the useful discussion on understanding the properties of AGNs. 
We are also grateful to Shasvath~J.~Kapadia and Aditya~Vijaykumar for their review and comments on the manuscript. 
We acknowledge the use of computing facilities supported by grants from the Croucher Innovation Award from the Croucher Foundation Hong Kong.
J.J. is supported by the research program of the Netherlands Organisation for Scientific Research (NWO).
P.M., S.H.W.L. and O.A.H. acknowledge support by grants from the Research Grants Council of Hong Kong (Project No.~CUHK~14304622 and 14307923), the start-up grant from The Chinese University of Hong Kong, and the Direct Grant for Research from the Research Committee of The Chinese University of Hong Kong. 
A.K.S. and P.A. acknowledge the support of the Department of Atomic Energy, Government of India, under project No. RTI4001.
\end{acknowledgments}
\software{
        NumPy~\citep{numpy},
        SciPy~\citep{scipy},
        Matplotlib \citep{matplotlib},
        Astropy~\citep{astropy:2013, astropy:2018, astropy:2022},
        Mathematica \citep{Mathematica},
        Cython~\citep{cython}
        }
\appendix

\section{Thin-disk geometry\label{sec:derivation}}
Throughout the text, we assume the source is confined to an infinitely thin disk, and a BBH on the disk can be represented by the following unit vector:
\begin{equation}
    \vu*S = \mqty(\cos\phi_S & \sin\phi_S & 0)\ ,
\end{equation}
and the optical axis is:
\begin{equation}
    \vu*O = -\vu*N = \mqty(-\sin\iota & 0 & -\cos\iota)\ .
\end{equation}
They form an angle $\gamma$:
 \begin{equation}
     \cos\gamma = \< \vu*S, \vu*O > = - \cos\phi_S\,\sin\iota\ ,
     \label{eq:gamma_relation}
\end{equation}
and from the geometry in Fig.~\ref{fig:agn_BBH_illustration}, the following relation holds:
\begin{equation}
    D_S\,\tan\beta = R_{\rm orb} \,\sin\gamma\ ,
    \label{eq:height_relation}
\end{equation}
where $\beta$ is the source angle. 
\new{In this appendix, we shall derive the expressions shown in Sec.~\ref{sec:AGN-BBH} with a relaxed condition, namely, we assume lensing will occur as long as $\beta / \theta_{\rm E} = y \leqslant y_{\rm max}$. In the main text, all results had $y_{\rm max}$ set to 1, for a more conservative constraint.}
% If lensing occurs, it is bounded by the Einstein radius:
\begin{align}
    \new{\frac{\beta}{y_{\rm max}}} \leqslant \theta_{\rm E} &= \sqrt{\frac{2R_{\rm Sch} D_{LS}}{D_S D_L}} \label{eq:beta_less_or_equal_to_Einstein_radius} \\
                             &\approx \frac{R_{\rm Sch}}{D_L}\sqrt{2 \frac{D_{LS}}{R_{\rm Sch}}}\ .
    \label{eq:beta_bound}
\end{align}
We made use of the fact that $D_S = D_L + D_{LS} \approx D_L$ on the second line. 
Applying this constraint to Eq.~\eqref{eq:height_relation}, we conclude
\begin{equation}
    2\,\cos\gamma \geqslant \new{\frac{r}{y_{\rm max}^2}}\, \sin^2\gamma
\end{equation}
where, as before, we defined $r = R_{\rm orb} / R_{\rm Sch}$, and assumed $D_S \approx D_L$ and the small angle approximation $\tan\beta \approx \beta$. Then, we express $\gamma$ in terms of  $\phi_S$ and  $\iota$ via Eq.~\eqref{eq:gamma_relation} as:
\begin{equation}
    \cos(\pi -\phi_S) \geqslant \frac{\new{y_{\rm max}^2}}{r\,\sin\iota}\qty(\sqrt{1 + \new{\qty(\frac{r}{y_{\rm max}^2})^2}} - 1)\ ,
    \label{eq:master_eqn}
\end{equation}
or equivalently, \new{if we write $\bar{r} = r / y_{\rm max}^2$},
\begin{equation}
    \phi_{S,{\rm min}} = \arccos( \frac{1 - \sqrt{1 + \new{\bar{r}}^2}}{\new{\bar{r}}\,\sin\iota})\ .
\end{equation}
This means that at a given $r$ and $\iota$, only BBHs that are within $ [\phi_{S,{\rm min}},\ 2\pi - \phi_{S,{\rm min}}]$ can be observed as lensed BBHs.
It translates into a probability:
\begin{align}
    p({\rm lensing} \mid r,\ \iota) 
    &= \frac{1}{2\,\pi} \qty[ (2\pi - \phi_{S,{\rm min}}) - \phi_{S,{\rm min}}] \\
    &= \frac{1}{\pi}\arccos( \frac{x}{\sin\iota})\ ,
    \tag{\ref{eq:conditional_prob}}
\end{align}
where $x$ was defined as
 \begin{equation}
     x = \frac{\sqrt{1 + \new{\bar{r}}^2} - 1}{\new{\bar{r}}}\ .
     \tag{\ref{eq:x_def}}
\end{equation}
Note that when $\iota = \pi /2$ (edge-on) and in the large $r$ limit, this probability drops as \new{$\sqrt{y_{\rm max}^2/r}$}.
However, due to the disk geometry, any deviation from the edge-on angle will result in a drastic drop in the probability of being lensed for binaries that are slightly farther away in the disk, as shown by the grey dashed line in Fig.~\ref{fig:lensed_prob}. 
We note that this grey line is the most conservative bound, for if the disk has finite height, this line becomes a grey band and extend the maximum radius more to the right.
At each given radius, Eq.~\eqref{eq:master_eqn} also defines a minimum inclination:
 \begin{equation}
     \iota_{\rm min} = \arcsin(x)\ ,
\end{equation}
we find that for $5\,R_{\rm Sch}$ (and beyond), no lensed binaries will be observed for inclination smaller than \SI{0.96}{\radian}.
From this we can define the \emph{probability of a lensed BBH in an AGN at a radius $r$} by marginalising over the observer angle:
\begin{align}
    f_{{\rm lensing}|{\rm AGN}} & = p({\rm lensing} \mid r) \nonumber \\
    & = \int_{\iota_{\rm min}}^{\pi /2} \frac{1}{\pi}\arccos( \frac{x}{\sin\iota}) \, \sin(\iota)\,\dd  \iota 
    \tag{\ref{eq:marg_prob}}\\
    & = \int_{1}^{x} \frac{1}{\pi}\arccos(y) \, \frac{1}{y^2} \frac{-x^2}{\sqrt{y^2 - x^2}} \,\dd y \\
    & = \frac{1 - x}{2} \label{eq:exact_f_lensing_AGN} 
\end{align}
the extra factor of $\sin(\iota)$ accounts for the isotropic probability distribution of the observer around the SMBH.
In the large $r$ limit, $x \to 1 - \new{y_{\rm max}^2}/r$, which gives us the approximation in Eq.~\eqref{eq:prob_approx} \new{when $y_{\rm max} = 1$; otherwise, we see that $f_{{\rm lensing}|{\rm AGN}} \propto y_{\rm max}^2$}.

\section{The AGN constraint\label{sec:constraint_dev}}

Assuming that we have $N_{\rm obs}$ BBHs, and a fraction $f_{{\rm lensed} \wedge {\rm AGN}}$ of them is both lensed and from an AGN, then the \emph{expected number of such lensed GW coming from an AGN} is the mean of its probability mass function, which is given by the binomial distribution:
\begin{equation}
    p(N_{{\rm lensed}} \mid N_{\rm obs}, k) = 
    {N_{\rm obs} \choose N_{\rm lensed} } \,
    k^{ N_{\rm lensed} } \,
    (1 - k) ^ { N_{\rm obs} - N_{\rm lensed}  }\ ,
    \label{eq:binomial_N}
\end{equation}
where $k = f_{{\rm lensed} \wedge {\rm AGN}}$ in this case. In order to find the upper bound of $f_{\rm AGN}$, we made the assumption that all lensed events originate from an AGN, hence we replace the $N_{{\rm lensed} \wedge {\rm AGN}}$ with simply $N_{{\rm lensed}} $. 
One can always relax this assumption with the knowledge of the proportions of lensed systems in our Universe.

Notice that $k = f_{{\rm lensed} \wedge {\rm AGN}}$ is an implicit function of $R_{\rm orb}$ and $f_{\rm AGN}$, and therefore we could write:
\begin{equation}
    p(N_{{\rm lensed}} \mid N_{\rm obs}, k) = p(N_{{\rm lensed}} \mid N_{\rm obs}, f_{\rm AGN}, R_{\rm orb})\ ,
\end{equation}
and from Bayes' theorem, we obtain:
\begin{equation}
    p(f_{\rm AGN} \mid N_{\rm obs}, N_{{\rm lensed}}) 
    \propto  \int p(N_{{\rm lensed}} \mid N_{\rm obs}, f_{\rm AGN}, R_{\rm orb})\, \pi(R_{\rm orb}) \, \dd R_{\rm orb}
    \label{eq:poste_f_AGN}
\end{equation}
and $\pi(R_{\rm orb})$ is the probability distribution of BBHs across the AGN disk, which we show in the lower panel of Fig.~\ref{fig:f_AGN_disk_constraint}. 
For the Gr\"ober~et.\,al. model, we adopted the radial part of Eq.~(14) in~\citet{grobner_binary_2020} as our radial distribution. 
For the SG~\citep{Sirko2003:AGNdisk} and TQM~\citep{Thompson2005:AGNdisk} models, we constructed the radial distribution from the density ($\rho(r)$) and aspect ratio ($h(r) / r$) profiles, as follow: $\pi(r) \propto r \times h(r) \times \rho(r)$.
Ideally, $\pi(r)$ should be proportional to the local merger rate of the disk, but if we assume all BBHs merges at the same rate, then the net rate is only proportional to the local density. Thus, it is sufficient to know the radial distribution of the binaries.

\section{Effects from the demagnified images}
For each pair of images that are lensed by the SMBH, one of them is always demagnified, and the amplitude of the signal will be reduced. 
Effectively, this means that the horizon distance of a given binary will be shortened.
In this section, we will estimate the maximum loss in merger rate due to the demagnified images.

First, for the unlensed signals, we model their merger rate as a function of the horizon redshift ($z_{\rm H}$) via:
\begin{equation}
    {\cal R}_{\rm U} (z_{\rm H}) = \int_{0}^{z_{\rm H}} \frac{R_{\rm Oguri}(z)}{1 + z} \dv{V_{\rm c}}{z} \, \dd z\ ,
    \label{eq:unlensed_rate}
\end{equation}
here, $\dd V_{\rm c} / \dd z$ is the differential comoving volume, and we assume the BBH distribution follows $R_{\rm Oguri}$, from~\citet{Oguri:2018muv}.

Then, for each binary, at a dimensionless position $y$ w.r.t. the SMBH lens, will receive a (de)magnification $\mu_\pm$ which depends on $y$ assuming the SMBH being a point mass lens.
Therefore, for each position $y$, the effective horizon distance is related to the unlensed horizon distance as follows~\citep{Allen:2005fk}:
\begin{equation}
    \frac{D_{\rm H}^{\rm eff}}{\sqrt{\mu_\pm(y)} \, D_{\rm H}} = \qty( \frac{1 + z(D_{\rm H}^{\rm eff})}{1 + z(D_{\rm H})} )^{5 / 6}\ ,
\end{equation}
note that the horizon redshift and distance are defined naturally as $z_{\rm H} = z(D_{\rm H})$, assuming the Planck-2018 cosmology.
Therefore, the magnified and demagnified merger rates can be computed by integrating Eq.~\eqref{eq:unlensed_rate} up to the effective horizon redshift for each $y$, 
which we further marginalised over by assuming  $p(y) \propto y$:
\begin{equation}
    {\cal R}_\pm (D_{\rm H}) = \int_{0}^{\new{y_{\rm max}}} p(y) \int_{0}^{z_{\rm H}^{\rm eff}(y)} \frac{R_{\rm Oguri}(z)}{1 + z} \dv{V_{\rm c}}{z} \, \dd z  \,\dd y .
\end{equation}
We found that for the horizon distance ranges from $ (100 \text{ to } 10^6)\,{\rm Mpc}$, the ratio between the demagnified merger rate and the unlensed one, $ {\cal R}_-(D_{\rm H}) / {\cal R}_{\rm U}(D_{\rm H})$, is about 90\% for $D_{\rm H} < 10^3\,\si{\Mpc}$ and is minimal at around \SI{3e4}{\Mpc} at $ {\sim}40\%$.
This means that for the LIGO binaries, it amounts to multiplying the lensing probability of Eq.~\eqref{eq:lensed_and_AGN} by 0.9, which would not significantly affect our results above, as the dominant source of uncertainty originates from the uncertainty in the orbital radii. 
On the other hand, with third-generation detectors, the horizon distance is far beyond \SI{3e4}{\Mpc}, in which case the ratio is close to unity, and there will be no bias resulting from demagnification.

\bibliography{AGN_references,LIGO_papers,bibliography}
\end{document}